\documentclass{elsart}

\usepackage{graphicx}
\usepackage{subfigure}
\usepackage{amsmath}

\input epsf

\begin{document}

\setcounter{footnote}{0}
\setcounter{figure}{0}

\begin{frontmatter}
\title{
Measurement of the $D^+$ and  $D_s^+$ decays into $K^+K^-K^+$. 
}

$\textrm{The~FOCUS~Collaboration}^\star$
\thanks{See \textrm{http://www-focus.fnal.gov/authors.html} for
additional author information.}

\author[ucd]{J.~M.~Link}
\author[ucd]{M.~Reyes}
\author[ucd]{P.~M.~Yager}
\author[cbpf]{J.~C.~Anjos}
\author[cbpf]{I.~Bediaga}
\author[cbpf]{C.~G\"obel}
\author[cbpf]{J.~Magnin}
\author[cbpf]{A.~Massafferri}
\author[cbpf]{J.~M.~de~Miranda}
\author[cbpf]{I.~M.~Pepe}
% ...

\address[ucd]{University of California, Davis, CA 95616}
\address[cbpf]{Centro Brasileiro de Pesquisas F\'isicas, Rio de Janeiro, RJ, Brasil}
\address[cinv]{CINVESTAV, 07000 M\'exico City, DF, Mexico}
\address[cu]{University of Colorado, Boulder, CO 80309}
\address[fnal]{Fermi National Accelerator Laboratory, Batavia, IL 60510}
\address[fras]{Laboratori Nazionali di Frascati dell'INFN, Frascati, Italy I-00044}
\address[ui]{University of Illinois, Urbana-Champaign, IL 61801}
\address[iu]{Indiana University, Bloomington, IN 47405}
\address[korea]{Korea University, Seoul, Korea 136-701}
\address[kp]{Kyungpook National University, Taegu, Korea 702-701}
\address[milan]{INFN and University of Milano, Milano, Italy}
\address[nc]{University of North Carolina, Asheville, NC 28804}
\address[pavia]{Dipartimento di Fisica Nucleare e Teorica and INFN, Pavia, Italy}
\address[pr]{University of Puerto Rico, Mayaguez, PR 00681}
\address[sc]{University of South Carolina, Columbia, SC 29208}
\address[ut]{University of Tennessee, Knoxville, TN 37996}
\address[vu]{Vanderbilt University, Nashville, TN 37235}
\address[wisc]{University of Wisconsin, Madison, WI 53706}

\begin{abstract}
We present the first clear observation of the doubly Cabibbo 
suppressed decay $D^+ \rightarrow K^-K^+K^+$ and the first observation of the 
singly Cabibbo suppressed decay $D_s^+ \rightarrow K^-K^+K^+$. 
These signals have been obtained by analyzing the high statistics sample of 
photoproduced charm particles of the FOCUS (E831) experiment at Fermilab. 
We measure the following relative branching ratios:
 
$\Gamma\left( D^+ \rightarrow K^- K^+ K^+ \right)  
/ \Gamma\left( D^+ \rightarrow K^- \pi^+ \pi^+ \right)= 
(9.49 \pm 2.17 \pm 0.22)\times 10^{-4}$

and

$\Gamma\left( D_s^+ \rightarrow K^- K^+ K^+ \right)/ \Gamma\left( D_s^+ 
\rightarrow K^- K^+ \pi^+ \right) = (8.95\pm 2.12~^{+2.24}_{-2.31})\times 10^{-3}$,
\\ \noindent
where the first error is statistical and the second is systematic.
\end{abstract}

\end{frontmatter}

\section{Introduction}

Doubly Cabibbo suppressed (DCS) charm decays are expected to occur with a rate 
which is roughly a factor $\tan^4\theta_C \sim 2.5 \times 10^{-3}$ smaller than 
the corresponding Cabibbo favored (CF) modes. This is the main reason 
our present knowledge of these decays is rather poor and limited 
to very few decay modes. Only four DCS decays have been observed, 
$D^+ \rightarrow K^+\pi^-\pi^+$, $D^0 \rightarrow K^+\pi^-$, 
$K^+\pi^-\pi^0$ and 
$K^+\pi^-\pi^+\pi^-$.\footnote{Evidence for the DCS decay
$D^+ \rightarrow K^- K^+ K^+$ was previously reported by two 
experiments~\cite{e691,wa82}, but their results were superseded~\cite{pdg} by 
the much more stringent upper limits coming from the higher statistic experiment 
E687~\cite{e687_limit}.} 
The interpretation of the $D^0$ modes is complicated by possible contributions
from $D^0-\overline{D^0}$ mixing~\cite{bergmann,jon,cleo,cleokpipi0}, 
making the $D^+ \rightarrow K^+\pi^-\pi^+$ decay the only pure DCS decay previously
studied.

In this paper, we report the first clear observation of  the DCS decay 
$D^+ \rightarrow K^-K^+K^+$, together with the first observation of the 
singly Cabibbo suppressed (SCS) decay of $D_s^+$ into the same final state. 
Throughout this paper, the charge conjugate 
is implied when a decay mode of a specific charge is stated.

It is interesting to note that in contrast to the four modes previously mentioned, 
the DCS decay $D^+ \rightarrow K^-K^+K^+$ cannot result from a simple spectator 
process, but presumably requires the intervention of strong resonances that 
simultaneously couple to the $\pi \pi$ and $KK$ channels. It could also proceed 
through annihilation but from studies of  $D_s^+ \rightarrow \pi^-\pi^+\pi^+$ 
we expect this contribution to be small~\cite{3pioni}.

The results presented in this paper have been obtained using the high statistics 
charm sample of the FOCUS experiment at Fermilab. 
FOCUS is a charm photoproduction experiment which took data during the 1996/1997 
fixed target run at Fermilab. 
The FOCUS detector is a large aperture, fixed-target spectrometer with
excellent vertexing and particle identification. A photon beam is
derived from the bremsstrahlung of secondary electrons and positrons
with an $\approx 300$ GeV endpoint energy produced from the 800~GeV/$c$ 
Tevatron proton beam. The photon beam interacts in a segmented BeO
target. The charged particles which emerge from
the target are tracked by two systems of silicon microvertex
detectors. The upstream system, consisting of 4 planes (two views in 2
stations), is interleaved with the experimental target, while the
other system lies downstream of the target and consists of twelve
planes of microstrips arranged in three views. These detectors provide
high resolution separation of primary (production) and secondary
(decay) vertices with an average proper time resolution of $\approx
30$~fs for 2-track vertices. The momentum of a charged particle
is determined by measuring its deflections in two analysis magnets of
opposite polarity with five stations of multiwire proportional
chambers. Three multicell threshold \v{C}erenkov counters are used to
discriminate between electrons, pions, kaons, and protons.

\section{Signals and selection criteria}

The final states are selected using a candidate driven vertex algorithm. 
The basic idea of this algorithm is to use a charm candidate decay vertex as a 
\emph{seed} to find the primary vertex. 
In our particular case a decay vertex is formed from three reconstructed charged tracks 
and the momentum vector of the resultant {\em D} candidate is used to 
intersect other reconstructed tracks and search for a suitable production vertex. 
The confidence levels of both vertices are required to be greater than $1\%$.
We measure $\ell$ the separation of the two vertices and its associated error
$\sigma_\ell$.
The quantity $\ell/\sigma_\ell$ is
the significance of detachment of the secondary and primary vertices.
Cuts on $\ell/\sigma_\ell$ are used to extract the $D$ signals from
non-charm background and to improve the signal to background ratio.
Two other measures of vertex isolation
are used: a {\em primary vertex isolation} and  a
{\em secondary vertex isolation}. The {\em primary vertex isolation} cut requires
that the confidence level for one of the tracks assigned to the decay 
vertex to be included in the primary vertex be less than a certain threshold value.
The {\em secondary vertex isolation} cut requires that the maximum confidence 
level for all tracks not assigned to any vertex to form a vertex with the $D$ 
candidate be less than a certain threshold value.
The main difference in the selection criteria between different decay modes 
lies in the particle identification cuts applied to the decay products. To 
minimize the systematic errors we use identical vertex cuts both on the signal 
and normalizing modes.

In the $D^+ \rightarrow K^-K^+K^+$ analysis we require $\ell/\sigma_\ell > 8$. The 
{\em primary} and {\em secondary vertex isolation} must be less than $0.1\%$. 
The $D$ momentum must be in the range $25~\textrm{GeV}/c$ to $250~\textrm{GeV}/c$ 
and the primary vertex 
must be formed with at least two reconstructed tracks in addition to the \emph{seed}
track. We require that the decay vertex occur outside of the target material.
For each charged track the \v{C}erenkov algorithm computes four 
likelihoods from the observed firing response of all the cells that 
lie inside the track's \v{C}erenkov cone for every counter~\cite{cerenkov}. 
The product of all firing probabilities for all cells within the three \v{C}erenkov
cones produces a $\chi^2$-like variable $W_i = -2 \ln (\mathrm{Likelihood})$, where
$i$ ranges over electron, pion, kaon and proton hypotheses. 
We require observed
\v{C}erenkov light pattern for the kaon hypothesis is favored over that for
the pion hypothesis by more than a factor of exp(0.5) by requiring
$W_{\pi} - W_K > 1.0$.
We also apply a kaon consistency cut, which requires that no
particle hypothesis is favored over the kaon hypothesis with a 
$\Delta W = W_K - W_{\textrm{min}}$ exceeding $3.5$. To further reduce 
the background due to poorly reconstructed candidates, we require 
that the proper time resolution of the candidates, defined as
$\sigma_\ell/(\beta\gamma c)$, be less than $ 150$~fs. 

The resulting $D^+$ signal is shown in Fig.1(a). We obtain a Gaussian yield of 
$65.5 \pm 15.0$ $D^+ \rightarrow K^-K^+K^+$ events over a linear background. The mass value returned 
by the fit is $1869 \pm 1~\textrm{MeV}/c^2$; the r.m.s. of the Gaussian fit is 
$5.2 \pm 1.2~\textrm{MeV}/c^2$ in agreement
with Monte Carlo simulations. 
The two broad structures around
$1985~\textrm{MeV}/c^2$ and $2085~\textrm{MeV}/c^2$ are  
due to $D^+$ and $D_s^+$ decays into $K^-K^+\pi^+$ where the $\pi^+$ is 
misidentified as a $K^+$. 

In the $D_s^+ \rightarrow K^-K^+K^+$ analysis we have to use stronger 
\v{C}erenkov cuts to extract the signal which otherwise would be completely hidden 
by the $K^-K^+\pi^+$ mis-identification peaks. We require $W_{\pi}-W_K > 4.5$ for all 
three kaon candidates. All the other cuts are the same as for
the $D^+ \rightarrow K^-K^+K^+$ decay.

Fig.1(b) shows the invariant mass plot where both $D^+$ and $D_s^+$ peaks are now evident.
In the fit the $D_s^+$ mass and width are fixed to the values found in the Monte Carlo.
This is done to reduce the effects of 
any residual fluctuation of the $D^+ \rightarrow K^-K^+\pi^+$ reflection, which would
induce a shift of the peak toward higher masses. We obtain a yield of $31.4 \pm 7.4$ 
$D_s^+ \rightarrow K^-K^+K^+$ events over a linear background.

For $D^+ \rightarrow K^-K^+K^+$ we measure the branching ratio relative to 
$D^+ \to K^-\pi^+\pi^+$, while for $D_s^+ \rightarrow K^-K^+K^+$ that relative to 
$D_s^+ \to K^-K^+\pi^+$. We obtain:

\begin{center}
$\Gamma\left( D^+ \rightarrow K^-K^+K^+ \right)/\Gamma\left( D^+ \rightarrow K^-
\pi^+ \pi^+ \right)
= (9.49 \pm 2.17)\times 10^{-4}$

\noindent
$\Gamma\left( D^+_s \rightarrow K^-K^+K^+ \right)/ \Gamma\left( D^+_s \rightarrow 
K^+ K^- \pi^+ \right) 
= (8.95\pm 2.12)\times 10^{-3}$.
\end{center}

The cuts on the normalization modes are identical whenever possible to those used 
for the selection of the corresponding $3K$ signal. 
In addition, 
to remove contamination from 
the $D_s^+ \to K^-K^+\pi^+$ normalization mode due to \v{C}erenkov misidentified 
$D^+ \to K^-\pi^+\pi^+$ events, we employ an {\it anti-reflection} 
cut to reject candidates which, when reconstructed as $K^-\pi^+\pi^+$, 
lie within $2$ sigma of the $D^+$ nominal mass.
The normalization
signals are shown in Fig.1(c) and Fig.1(d) and consist of $62911 \pm 263$ 
and $3844 \pm 66$ events respectively.

In all our simulations we always used the proper resonant substructure for the 
two normalization modes~\cite{e687_kpp}~\cite{e687_kkp}, which would otherwise produce important systematic deviations 
of the results.

\setcounter{figure}{0}

\begin{figure}[!htp]
\centering
\subfigure[]
{\includegraphics[width=0.4\textwidth,clip=true]{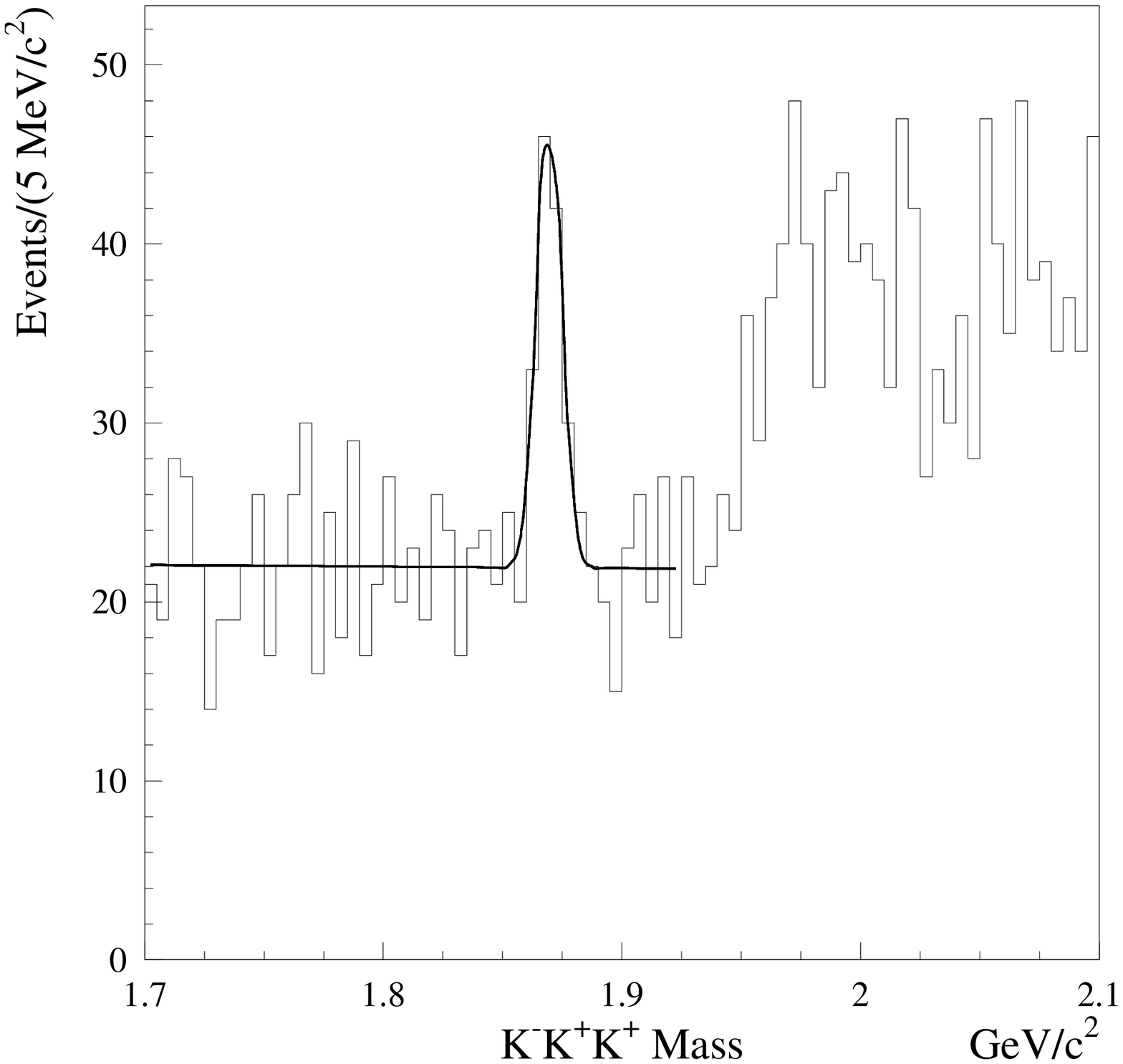}}
\subfigure[]
{\includegraphics[width=0.4\textwidth,clip=true]{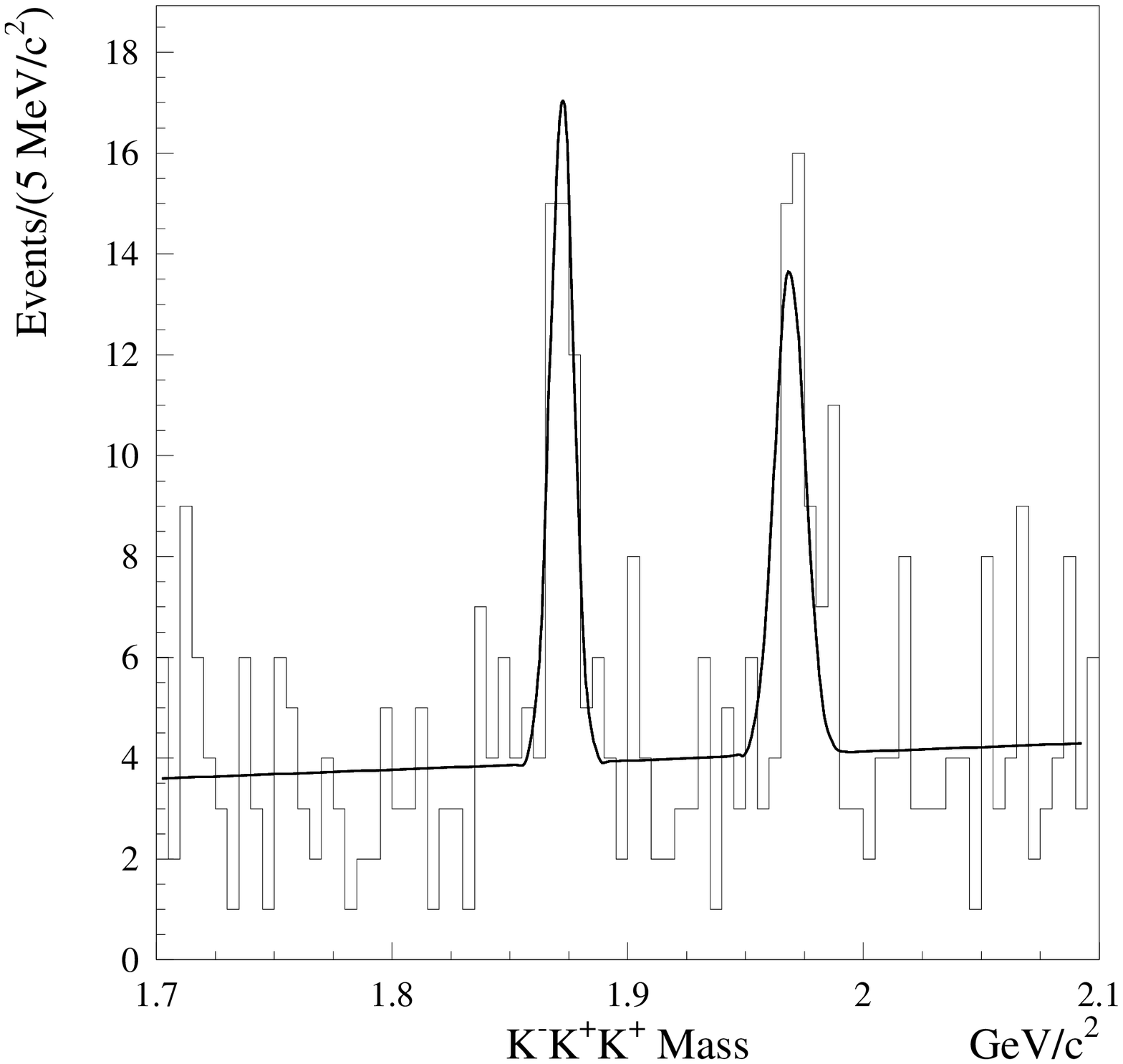}} \\
\subfigure[]
{\includegraphics[width=0.4\textwidth,clip=true]{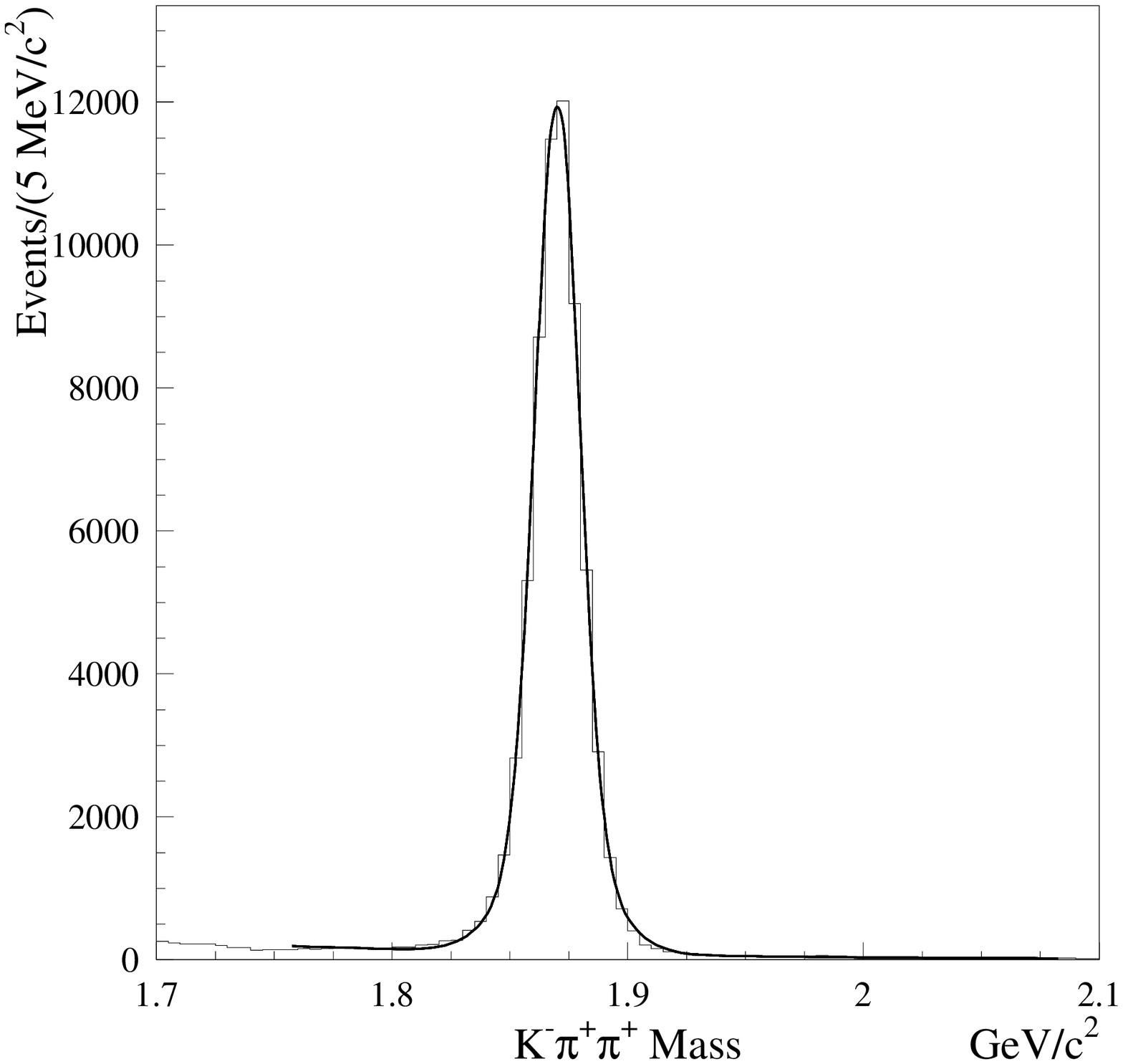}}
\subfigure[]
{\includegraphics[width=0.4\textwidth,clip=true]{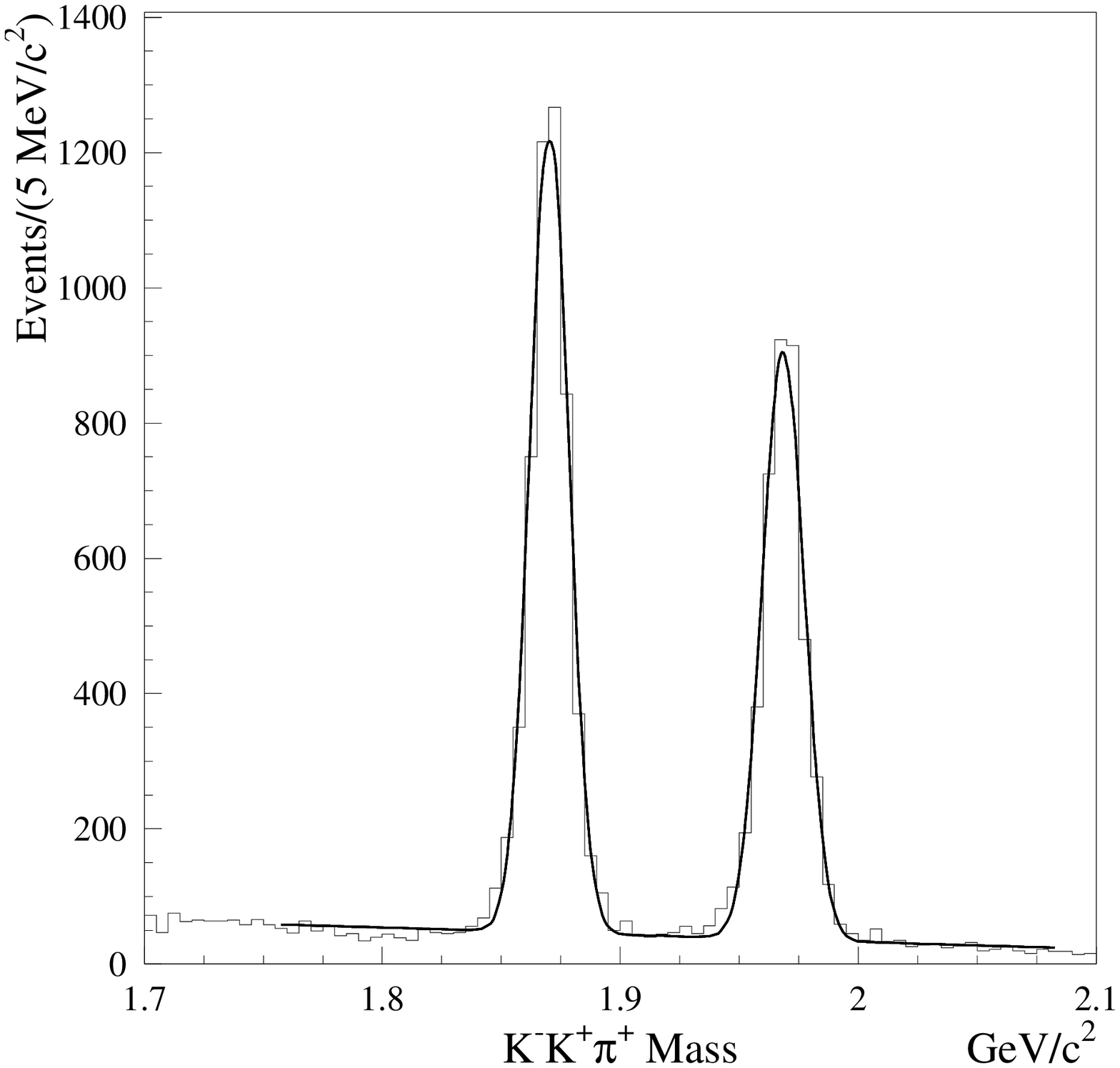}}
\label{masses}
\caption{Invariant mass distributions for $D^+ \to K^-K^+K^+$(a), $D_s^+ \to K^-K^+K^+$(b), 
$D^+ \to K^-\pi^+\pi^+$(c) and $D_s^+ \to K^-K^+\pi^+$(d).}
\end{figure}

\section{Systematic Errors}

We performed a detailed investigation of any source of 
systematics which could impact our branching ratio measurements. 
We first studied the stability of the results by varying the cuts over a wide range 
of values.
Our results are stable in their evolution on the most critical cuts:
$\ell/\sigma_\ell$, $W_{\pi}-W_K$ and  {\em primary} and 
{\em secondary vertex isolation}.
 
We then split the samples using variables which can probe different kinematical 
regions, such as low and high momentum range, or different experimental conditions, 
such as early and late runs, which have different target configurations. 
In doing this we can check our results together with our Monte Carlo 
simulation over a variety of different conditions.
We quantify a ``split sample systematic error" by examining consistency among 
these statistically independent splits of our data.
If the consistency $\chi^2$ turns out to be smaller than $1$, this error 
is taken to be zero. Otherwise 
we scale all the errors up to bring the $\chi^2$ back 
to $1$. The split sample systematic error is then defined as the difference in quadrature 
between the scaled error of the weighted average of the subsample estimates and 
the statistical error of the total data set.
This procedure is similar to the $S$-factor method used by the Particle Data Group~\cite{pdg}.

We have split our sample by high and low $D$-momentum, $D$ and $\bar{D}$, and 
early and late run periods.
Splits have been done in one variable at a time because of our limited 
statistics.

The measured branching ratios for the three pairs of disjoint 
samples are shown in Fig.2. 
We find only one contribution to the systematic uncertainty, namely the run-period split
sample for the $D_s^+$ decay
which gives 
a contribution to the branching ratio systematics of 
2.23$\times$10$^{-3}$.
 
\begin{figure}[!htp]
\centering
\subfigure[]
{\includegraphics[width=0.4\textwidth,clip=true]{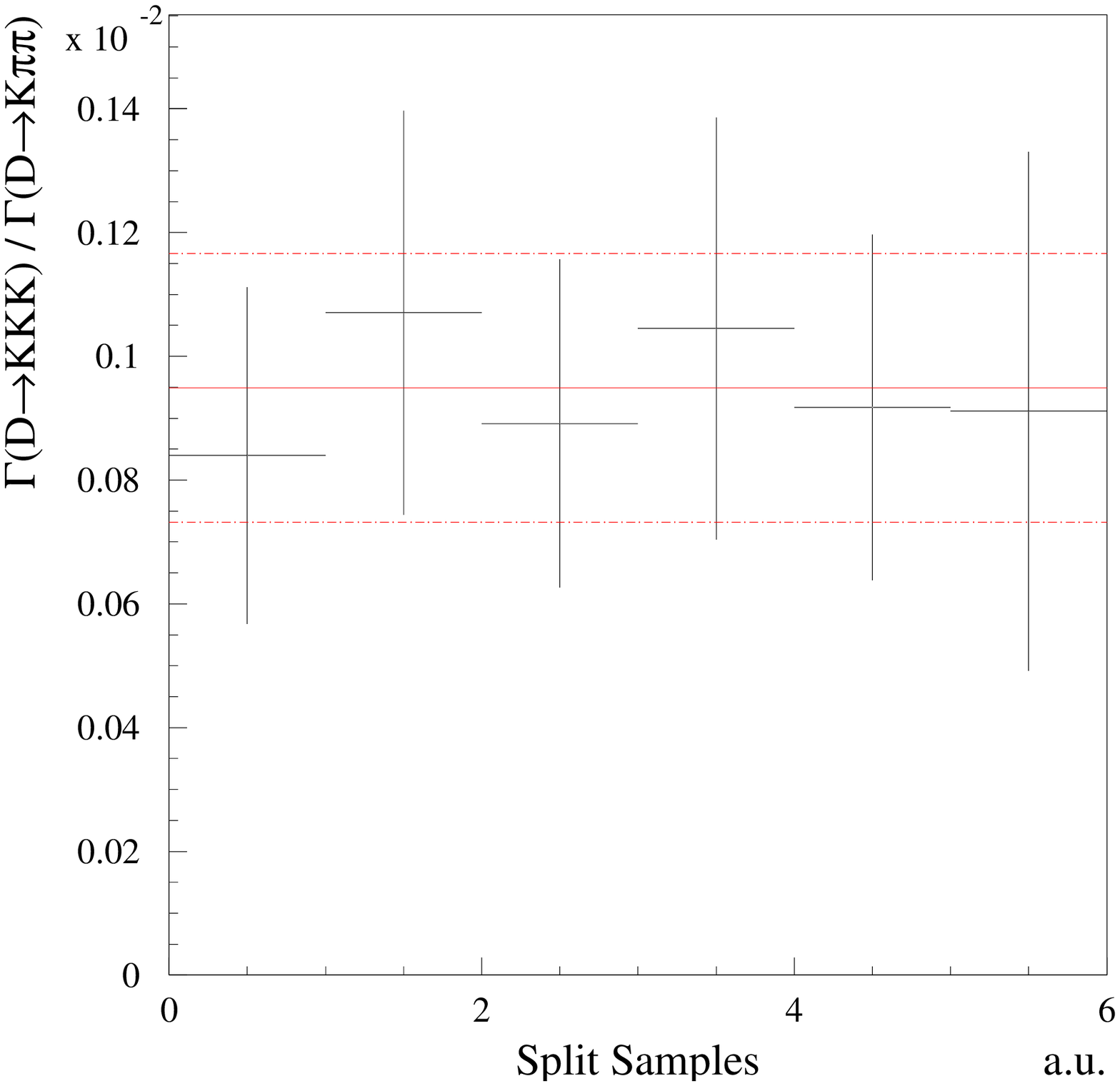}}
\subfigure[]
{\includegraphics[width=0.4\textwidth,clip=true]{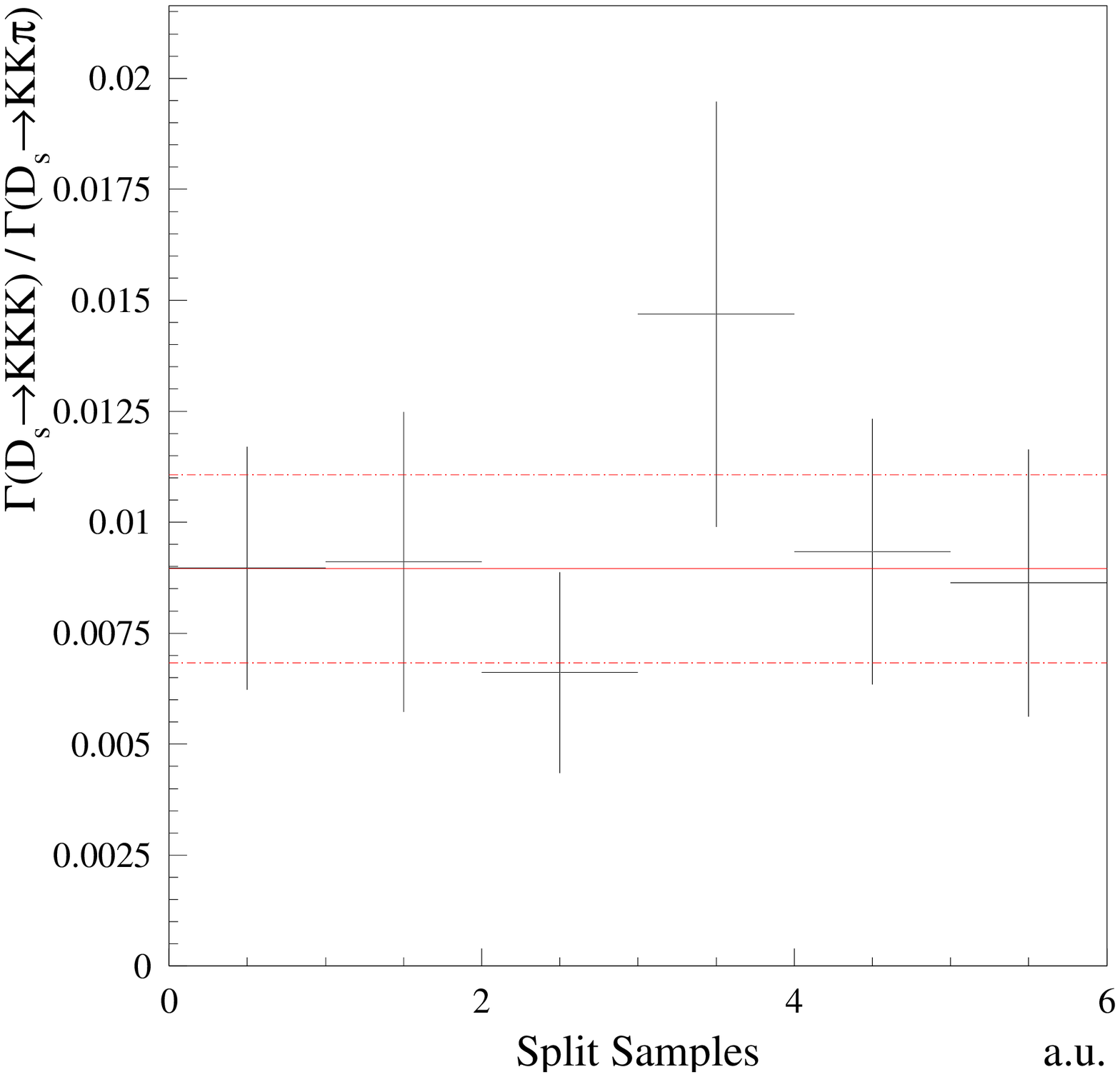}}
\label{splits}
\caption{Split sample results for $D$(a) and $D_s$(b) relative branching ratios.
Three pairs of disjoint samples are considered: high and low momenta on the left,
late and early runs in the center, $D$ and $\bar{D}$ on the right. The lines show
the joint sample and the $1 \sigma$ error bars. }
\end{figure}

In computing the branching ratios we have used the efficiency of a pure phase-space 
decay.
This choice was motivated by the relatively flat distribution of the events over the
Dalitz domains as shown in Fig. 3.
\begin{figure}[!htp]
\centering
\subfigure[]
{\includegraphics[width=0.4\textwidth,clip=true]{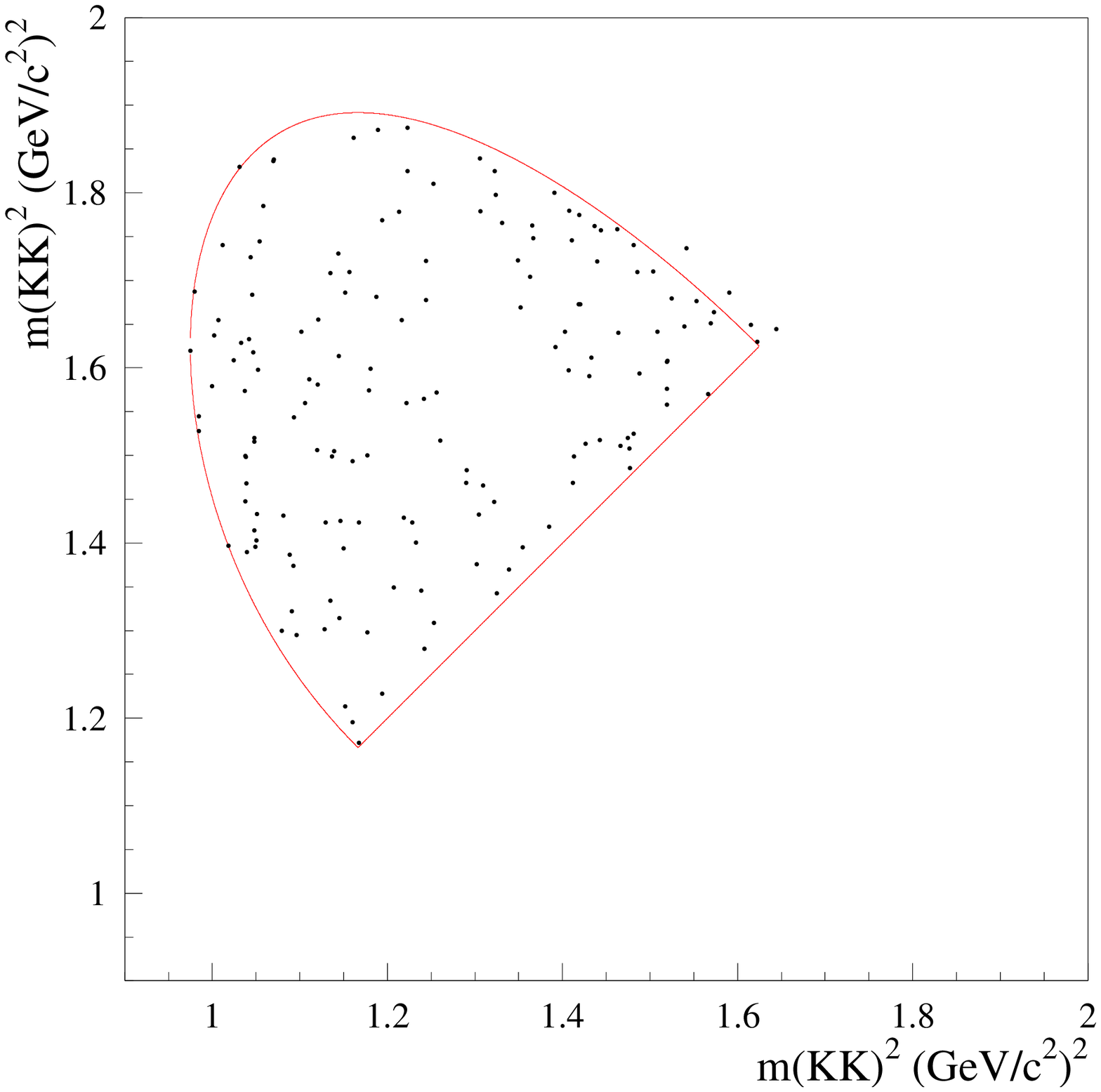}}
\subfigure[]
{\includegraphics[width=0.4\textwidth,clip=true]{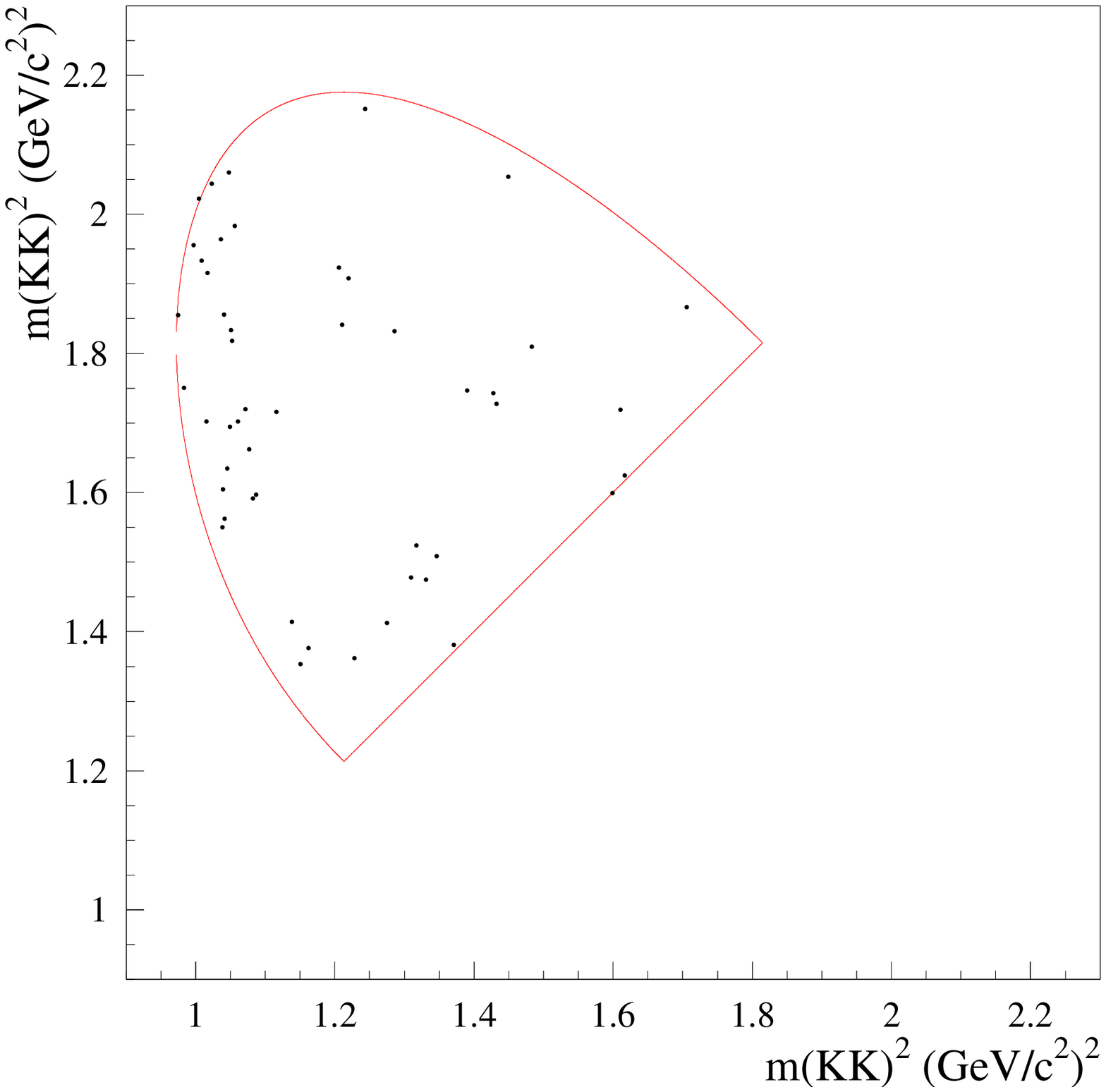}}
\label{fig:dalitz_plots}
\caption{Dalitz plot for $D^+$(a) and for $D_s^+$(b).
Only events which lie within $2 \sigma$ of the respective nominal masses are plotted.}
\end{figure}
To better investigate the implications of this assumption we have computed the 
reconstruction efficiencies for two particularly representative cases, 
a $\phi K^+$ decay and 
a $f_0(980) K^+$ decay. Table~\ref{tab:efficienze} shows the calculated efficiencies 
with respect to those for pure phase-space decays.
\begin{table}[!htb]
\begin{center}
\begin{tabular}{|c|c|c|} \hline
			& $\epsilon (D^+)$	& $\epsilon (D_s^+)$   \\ \hline \hline
      Phase-Space	& $1$			& $1$				\\ \hline
        $\phi K^+$	& $0.927\pm 0.015$	& $0.948\pm 0.015$		\\ \hline
       $f_0(980) K^+$	& $1.028\pm 0.014$	& $1.086\pm 0.014$		\\ \hline
\end{tabular}
\caption{Reconstruction efficiencies, $\epsilon$, for different decay dynamics into the 
same $K^-K^+K^+$ final state for $D^+$ and $D_s^+$.}
\label{tab:efficienze}
\end{center}
\end{table}
Given the non-negligible variation of the efficiency values, we 
considered the following two cases 
in order to assess the systematic uncertainty: the decay proceeds through the maximum 
estimated amount of $\phi K^+$ component, the remaining being pure phase space; 
the decay proceeds through the maximum estimated amount of $f_0(980) K^+$ component, 
the remaining being pure phase space. The estimated fractions, 
shown in Table~\ref{tab:reflex}, have been obtained by fits to the 
$3K$ invariant mass plots requiring that the $K^+K^-$ invariant mass lie within 
$2 \sigma$ of the nominal $\phi$ mass for the $\phi K^+$ decay and between two kaon mass 
threshold and 
$1.05~\textrm{GeV/}c^2$ for the $f_0(980) K^+$ decay. These estimates are crude and represent 
conservative upper limits for the purpose of estimating systematic errors and
are not meant to be measurements.\footnote{We consider these as conservative 
upper limits since we do not account for the contribution of other components below the 
$\phi$ and, when quoting the $f_0(980)K^+$ fraction, we do not 
simultaneously account for the $\phi$.}
\begin{table}[!htb]
\begin{center}
\begin{tabular}{|c|c|c|} \hline
			& $D^+$	                & $D_s^+$                \\ \hline 
      $\phi K^+$	& $12.4\%$		& $18.75\%$		\\ \hline
       $f_0(980) K^+$	& $44.5\%$		& $72\%$		\\ \hline
\end{tabular}
\caption{Estimated fraction of $\phi K^+$ and $f_0(980) K^+$ components for $D^+$ and
$D_s^+$ decays.}
\label{tab:reflex}
\end{center}
\end{table}
Under these assumptions, the contribution to the total systematics on the branching ratio 
measurement is $\pm 0.10\times 10^{-4}$ 
for $D^+$ and $^{+0.09}_{-0.52}\times 10^{-3}$ for $D_s^+$. 

The last source of systematic error we studied is that due to fitting procedure.
We calculated our branching ratios for various fit conditions, such as
changing the parametrization of the background shapes, rebinning the histograms, including in 
the $D^+$ fit the $K^- K^+ \pi^+$ reflection peaks and varying 
the fixed $D_s^+$ mass value by $1 \sigma$ of the quoted error~\cite{pdg}.
Since all these results are a priori likely we used the resulting sample variance to 
estimate the associated systematics. We obtain a systematic contribution
of $\pm$0.19$\times$10$^{-4}$
for the $D^+$ decay mode and $^{+0.12}_{-0.33}\times 10^{-3}$ for the $D_s^+$.

In conclusion, summing in quadrature the different systematic errors we obtain our final results:

\begin{center}
$BR\left( D^+ \rightarrow K^-K^+K^+ \right)/\left( D^+ \rightarrow 
K^-\pi^+\pi^+ \right)
= (9.49 \pm 2.17 \pm 0.22) \times 10^{-4}$
\end{center}

and

\begin{center}
$BR\left( D^+_s \rightarrow K^-K^+K^+ \right)/ \left( D^+_s \rightarrow 
K^-K^+\pi^+ \right) =
(8.95\pm 2.12~^{+2.24}_{-2.31})\times 10^{-3}$
\end{center}
\vspace{1.cm}

\section{Conclusions}

Our $D^+$ measurement is consistent with the
E687 upper limit~\cite{e687_limit} and constitutes the first clear evidence 
for this DCS decay. 
Our data indicate that only a minor fraction, if any, of the decay proceeds through the
$\phi K^+$ channel. 
This could suggest that the
decay proceeds mainly through resonances that can couple to both $\pi \pi$ and $KK$, such as 
the $f_0$ resonance series, as expected from a naive spectator picture.
However, more statistics would be needed to make quantitative statements through
 a Dalitz analysis.

Our $D_s^+$ measurement is consistent with the
E687 upper limit~\cite{e687_limit} and represents the first observation of the 
$3K$ mode. It constitutes the second Cabibbo suppressed decay of the $D_s^+$ 
measured. 
For Cabibbo suppressed decays other than 
$D_s^+ \rightarrow K^+ \pi^- \pi^+$~\cite{e687_kpp_cs}, 
only upper limits exist~\cite{mark3_k0pi}.

\section{Acknowledgments}
We wish to acknowledge the assistance of the staffs of Fermi National
Accelerator Laboratory, the INFN of Italy, and the physics departments of
the
collaborating institutions. This research was supported in part by the U.~S.
National Science Foundation, the U.~S. Department of Energy, the Italian
Istituto Nazionale di Fisica Nucleare and Ministero dell'Istruzione 
dell'Universit\`a e della Ricerca, the Brazilian Conselho Nacional de
Desenvolvimento Cient\'{\i}fico e Tecnol\'ogico, CONACyT-M\'exico, the
Korean
Ministry of Education, and the Korean Science and Engineering Foundation.

%
%
%============================  REFERENCES =======================
%
%***** papers and notes

\end{document}